\documentclass[prl,amssymb,twocolumn,showpacs,superscriptaddress,floatfix]{revtex4}

\usepackage{graphicx}
\usepackage{bm}
\usepackage{mathptmx}

\begin{document}

\title{Critical exponents and universality for the isotropic-nematic phase transition in a
system of self-assembled rigid rods on a lattice}

\author{L. G. L\'opez}
\author{D. H. Linares}
\author{A. J. Ramirez-Pastor}
\email{antorami@unsl.edu.ar} \affiliation{Departamento de
F\'{\i}sica, Instituto de F\'{\i}sica Aplicada, CONICET,
Universidad Nacional de San Luis,  5700 San Luis, Argentina}

\begin{abstract}
Monte Carlo simulations have been carried out for a system of
monomers on square lattices that, by decreasing temperature or
increasing density, polymerize reversibly into chains with two
allowed directions and, at the same time, undergo a continuous
isotropic-nematic (IN) transition. The results show that the
self-assembly process affects the nature of the transition. Thus,
the calculation of the critical exponents and the behavior of
Binder cumulants indicate that the universality class of the IN
transition changes from two-dimensional Ising-type for
monodisperse rods without self-assembly to $q=1$ Potts-type for
self-assembled rods.
\end{abstract}

 %Numerical simulation studies
\date{\today}
\maketitle

Self-assembly is a challenging field of research, driven
principally by the desire to design new materials. Moreover,
self-assembly is used permanently in biological systems to
construct supramolecular structures such as virus capsids,
filaments, and many others large molecular complexes. So,
understanding the rules of self-assembly has important
applications to both materials science and biology \cite{Service}.

On the other hand, the isotropic-nematic (IN) transition in
solutions of rodlike particles has been attracting a great deal of
interest since long ago. A seminal contribution to this subject
was made by Onsager \cite{ONSAGER} with his paper on the IN
transitions of infinitely thin rods. This theory shows that
particles interacting with only excluded volume interaction may
exhibit a rich phase diagram, despite the absence of any
attraction. Later, computer simulations of hard ellipses of finite
length \cite{BARON} confirmed the Onsager prediction that particle
sha-pe anisotropy can be a sufficient condition to induce the
long-range orientational order found in nematic liquid crystals.

In contrast to ordinary liquid crystal, many rodlike biological
polymers are formed by monomers reversibly self-assembling into
chains of arbitrary length so that these systems exhibit a broad
equilibrium distribution of filament lengths. An experimental
contribution to the study of these systems has been presented by
Viamontes et al. \cite{VIAMONTES}. The authors reported a
continuous IN transition for solutions of long F-actin (average
filament length longer than 2$\mu$m) and showed the existence of a
first-order phase separation for solutions of F-actin with average
filament length shorter than 2$\mu$m. These findings contradict
what is generally accepted in the literature: in three dimensions,
the IN transition is typically first order. On the other hand, in
two dimensions both continuous \cite{Frenkel} and first-order
\cite{Vink0} IN transitions can occur. Here, we consider a
self-assembled two-dimensional (2D) system that undergo a IN
transition, which is expected to be a continuous phase transition
\cite{Tavares}.

As mentioned above, the self-assembled system is intrinsically
polydisperse. While being able to solve explicitly only the
monodisperse case, Onsager \cite{ONSAGER} already outlined the
possible extension of the theory to polydisperse systems. In this
line of work, a detailed investigation of the effects of full
length polydispersity, i.e., of a continuous distribution of rod
lengths, on the Onsager theory has been recently developed by
Speranza and Sollich \cite{SPERA}. Another approach to the problem
of monodisperse rodlike mixtures has been proposed by Zwanzig
\cite{ZWA}. The Zwanzig model has been also extended to
polydisperse systems \cite{Clarke}, providing thus a useful
starting point for understanding the effects of polydispersity on
the phase behavior of hard rod systems. However, a complete
description of a system of self-assembled rods should consider not
only the effects of polydispersity, but also the influence of the
polymerization process.

In this context, we focus on a system composed of monomers with
two attractive (sticky) poles that polymerize reversibly into
polydisperse chains and, at the same time, undergo a continuous
phase transition. So, the interplay between the self-assembly
process and the nematic ordering is a distinctive characteristic
of these systems.

The same system has been recently considered by Tavares et al.
\cite{Tavares}. Using an approach in the spirit of the Zwanzig
model, the authors studied the IN transition occurring in a
two-dimensional system of self-assembled rigid rods. The obtained
results revealed that nematic ordering enhances bonding. In
addition, the average rod length was described quantitatively in
both phases, while the location of the ordering transition, which
was found to be continuous, was predicted semiquantitatively by
the theory.

Despite these interesting results there is an open question to be
answered: ``what type of phase transition is it?" Tavares et al.
\cite{Tavares} assumed as working hypothesis that the nature of
the IN transition remains unchanged with respect to the case of
monodisperse rigid rods on square lattices, where the transition
is in the 2D Ising universality class \cite{Ghosh,EPL1}. In this
context, the confirmation (or not) of this hypothesis is not only
important to resolve the universality class of the IN transition
occurring in a system of self-assembled rods, but also to shed
light on our understanding of the effect of the self-assembly
process on the nature of the transition. The objective of this
Rapid Communication is to provide a thorough study in this
direction. For this purpose, extensive Monte Carlo (MC)
simulations supplemented by analysis using finite-size scaling
(FSS) theory \cite{BINDER} have been carried out to study the
critical behavior in a system of self-assembled rigid rods
deposited on square lattices with two allowed directions. The
calculations were developed at constant temperature and different
densities, thus allowing a direct comparison with previous results
for long monodisperse rigid rods on two-dimensional lattices
\cite{Ghosh,EPL1}. Then, the conventional normalized scaling
variable $\epsilon \equiv T/T_c - 1$ was replaced by $\epsilon
\equiv \theta/\theta_c - 1$, where $T$, $T_c$, $\theta$, and
$\theta_c$ represent temperature, critical temperature, density,
and critical density, respectively. A nematic phase, characterized
by a big domain of parallel self-assembled rigid rods, is
separated from the disordered state by a continuous IN transition
occurring at a finite critical density. The results show that the
self-assembly process affects the nature of the transition. Thus,
the determination of the critical exponents indicates that the
universality class of the IN transition changes from 2D Ising-type
for monodisperse rods without self-assembly \cite{EPL1} to $q=1$
Potts-type for self-assembled rods.

In a recent paper, Fischer and Vink \cite{FISCHER} indicated that
the transition studied in Refs. \cite{Ghosh,EPL1} corresponds to a
liquid-gas transition rather than IN. This interpretation is
consistent with the 2D-Ising critical behavior observed for
monodisperse rigid rods on square lattices. However, as mentioned
in the previous paragraph, the universality class of
self-assembled rods changes from that of the 2D Ising model and
the result in Ref. \cite{FISCHER} is not generalizable to the
system studied here. Accordingly, we will continue using the term
``IN phase transition" as in the previous work of Tavares et al.
\cite{Tavares}.

%%%%%% MODEL and PARAMETER %%%%%

As in Ref. \cite{Tavares}, we consider a system of self-assembled
rods with a discrete number of orientations in 2D. We assume that
the substrate is represented by a square lattice of $M=L \times L$
sites with periodic boundary conditions. $N$ particles are
adsorbed on the substrate with two possible orientations along the
principal axis of the lattice. These particles interact with
nearest neighbors (NN) through anisotropic attractive
interactions. Then the adsorbed phase is characterized by the
Hamiltonian $H =  \sum_{\langle i,j \rangle} w_{ij} c_i c_j$,
where $\langle i,j \rangle$ indicates a sum over NN sites;
$w_{ij}$ represents the NN lateral interaction, which is
$w_{ij}=-w$ if two neighboring particles $i$ and $j$ are aligned
with each other and with the intermolecular vector, and is
$w_{ij}=0$ otherwise; and $c_i$ is the occupation variable with
$c_i=0$ if the site $i$ is empty, and $c_i=1$ if the site $i$ is
occupied.

%%%%%%%%%%%%%%%%%%%%%%%%%%%%%%%%%%

\begin{figure}
\includegraphics[width=5.40cm,clip=true]{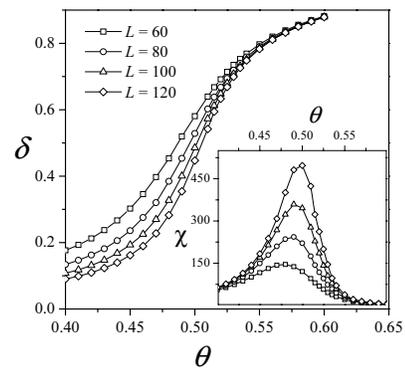}
\caption{\label{figure1} Size dependence of the order parameter
($\delta$) as a function of density ($\theta$). Inset: size
dependence of the susceptibility ($\chi$) as a function of density
($\theta$).}
\end{figure}

\begin{figure}
\includegraphics[width=5.40cm,clip=true]{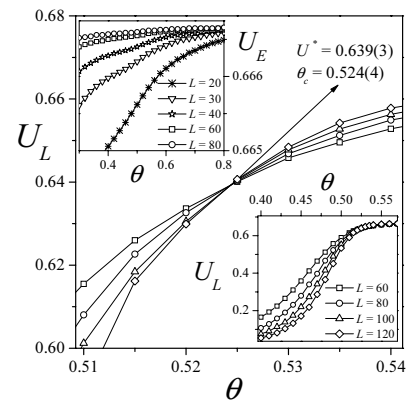}
\caption{\label{figure2} Curves of $U_L(\theta)$ vs $\theta$ for
square lattices of different sizes. From their intersections one
obtained $\theta_c$. In the inset, the data are plotted over a
wider range of densities.}
\end{figure}

A cluster or uninterrupted sequence of bonded particles is a
self-assembled rod. At fixed temperature, the average rod length
increases as the density increases and the polydisperse rods will
undergo a nematic ordering transition \cite{Tavares}. In order to
follow the formation of the nematic phase from the isotropic
phase, we use the order parameter defined in Ref. \cite{Tavares},
which can be written as $\delta =  \left | N_h -N_v \right
|/\left( N_h +N_v \right)$, where $N_h(N_v)$ is the number of
particles in clusters aligned along the horizontal (vertical)
direction. When the system is disordered, all orientations are
equivalents and $\delta$ is zero. In the critical regime, the
particles align along one direction and $\delta$ is different from
zero. In other words, the IN phase transition is accomplished by a
breaking of the orientational symmetry and $\delta$ appears as a
proper order parameter to elucidate this phenomenon.

The problem has been studied by canonical Monte Carlo simulations
using an vacancy-particle-exchange Kawasaki dynamics \cite{Kawa}
and Metropolis acceptance probability \cite{Metropolis}.
Typically, the equilibrium state can be well reproduced after
discarding the first $5 \times 10^6$ Monte Carlo steps (MCS).
Then, the next $6 \times 10^8$ MCS are used to compute averages.

In our Monte Carlo simulations, we set the temperature $T$, varied
the density $\theta=N/M$, and monitored the order parameter
$\delta$, which can be calculated as simple average. The
quantities related with the order parameter, such as the
susceptibility $\chi$, and the reduced fourth-order cumulant $U_L$
introduced by Binder~\cite{BINDER} were calculated as $\chi =
\frac{L^2}{k_BT} [ \langle \delta^2 \rangle - \langle \delta
\rangle^2]$ and $U_L = 1 -\langle \delta^4\rangle /\left[ 3\langle
\delta^2\rangle^2 \right]$, where $\langle \cdots \rangle$ means
the average over the MC simulation runs. In addition, in order to
discuss the nature of the phase transition, the fourth-order
energy cumulant $U_E$ was obtained as $U_E = 1 -\langle H^4\rangle
/\left[ 3\langle H^2\rangle^2 \right]$.

%%%%%% RESULTS %%%%%

\begin{figure}
\includegraphics[width=5.40cm,clip=true]{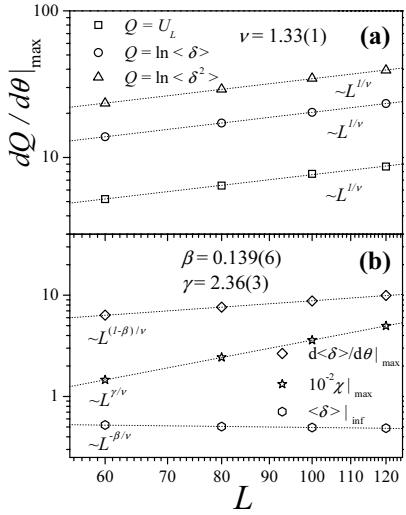}
\caption{\label{figure3} (a) Log-log plot of the size dependence
of the maximum values of derivatives of various thermodynamic
quantities used to determine $\nu$.  (b) Log-log plot of the size
dependence of the maximum value of the susceptibility, the point
of inflection of the order parameter and the maximum value of the
derivative of the order parameter used to determine $\gamma$ and
$\beta$, respectively.}
\end{figure}

\begin{figure}
\includegraphics[width=5.40cm,clip=true]{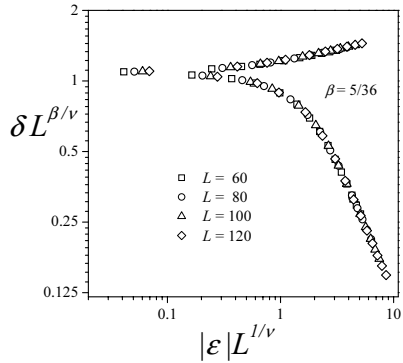}
\caption{\label{figure4} Data collapsing of the order parameter,
$\delta L^{\beta/\nu}$ vs $|\epsilon| L^{1/\nu}$. The plot was
made using $\theta_c=0.524$ and the exact percolation exponents
$\nu=4/3$ and $\beta=5/36$.}
\end{figure}

\begin{figure}
\includegraphics[width=5.40cm,clip=true]{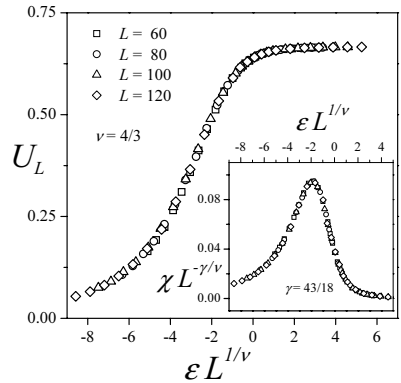}
\caption{\label{figure5} Data collapsing of the cumulant, $U_L$ vs
$\epsilon L^{1/\nu}$, and of the susceptibility, $\chi
L^{-\gamma/\nu}$ vs $\epsilon L^{1/\nu}$ (inset). The plots were
made using $\theta_c=0.524$ and exact percolation exponents
$\nu=4/3$ and $\beta=43/18$.}
\end{figure}

The critical behavior of the present model has been investigated
by means of the computational scheme described in the previous
paragraphs and finite-size scaling analysis. The FSS theory
implies the following behavior of $\delta$, $\chi$, and $U_L$ at
criticality: $\delta = L^{-\beta/\nu} \tilde \delta(L^{1/\nu}
\epsilon)$, $\chi= L^{\gamma/\nu}\tilde \chi(L^{1/\nu} \epsilon)$,
and $U_L=\tilde U_L(L^{1/\nu} \epsilon)$ for $L \rightarrow
\infty$, $\epsilon \rightarrow 0$ such that $L^{1/\nu} \epsilon $=
finite, where $\epsilon \equiv \theta/\theta_c - 1$. Here $\beta$,
$\gamma$, and $\nu$ are the standard critical exponents of the
order parameter, susceptibility, and correlation length,
respectively. $\tilde \delta, \tilde \chi $ and $\tilde U_L$ are
scaling functions for the respective quantities.

The phase diagram of the system under study has been recently
reported by Tavares et al. \cite{Tavares}. The authors showed that
the critical density, at which the IN transition occurs, increases
monotonically as $k_BT/w$ is increased. Thus, the nematic phase is
stable at low temperatures and high densities [see Fig. 1(a) in
Ref. \cite{Tavares}]. In addition, Tavares et al. found strong
numerical evidence that the IN transition is a continuous phase
transition. However, the authors were not able to determine the
critical quantities characterizing the universality class of the
mentioned transition. In the following, we try to resolve this
problem.

In our study and based on the phase diagram given in Ref.
\cite{Tavares}, we set the lateral interaction to $w=4k_BT$. With
this value of $w$, it is expected the appearance of a nematic
phase at intermediate densities. Accordingly, the density was
varied between $0.4$ and $0.6$. For each value of $\theta$, the
effect of finite size was investigated by examining square
lattices with $L= 60, 80, 100$, and $120$.

%%%%%% FIGURA 1 y 2%%%%%

We start with the calculation of the order parameter (Fig. 1),
susceptibility (inset in Fig. 1), and cumulant (Fig. 2) plotted
versus $\theta$ for several lattice sizes. In the vicinity of the
critical point, cumulants show a strong dependence on the system
size. However, at the critical point the cumulants adopt a
nontrivial value $U^*$; irrespective of system sizes in the
scaling limit. Thus, plotting $U_L(\theta)$ for different linear
dimensions yields an intersection point $U^*$, which gives an
accurate estimation of the critical density in the infinite system
and allows us to make a preliminary identification of the
universality class of the transition \cite{BINDER}. In this case,
the values obtained for the critical density and the intersection
point of the cumulants were $\theta_c=0.524(4)$ and $U^*
=0.639(3)$, respectively. This fixed value of the cumulants has
changed from that obtained for monodisperse rigid rods on square
lattices [$U^* =0.615(5)$], which may be taken as a first
indication that the universality class of the present model is
different from the well-known 2D Ising-type for monodisperse rods
\cite{EPL1}. In the lower-right inset, the data are plotted over a
wider range of densities. As can be seen, the curves exhibit the
typical behavior of $U_L$ in the presence of a continuous phase
transition. Namely, the order-parameter cumulant shows a smooth
increase from 0 to 2/3 instead of the characteristic deep
(negative) minimum, as in a first-order phase transition
\cite{BINDER}.

In order to discard the possibility that the phase transition is a
first-order one, the energy cumulants have been measured for
different lattice sizes ranging between $L = 20$ and $L = 80$. As
is well known, the finite-size analysis of $U_E$ is a simple and
direct way to determine the order of a phase transition
\cite{BINDER}. Our results for $U_E$ show a dip close to the
critical density $\theta_c= 0.524$ for all system sizes, but this
minimum scales to 2/3 in the thermodynamic limit as can be seen in
the upper-left inset of Fig. 2. These results exclude a
first-order transition, confirming the predictions by Tavares et
al. \cite{Tavares}.

%%%%%% FIGURA 3 %%%%%

Next, the critical exponents will be calculated. As stated in Ref.
\cite{Ferren}, the critical exponent $\nu$ can be obtained by
considering the scaling behavior of certain thermodynamic
derivatives with respect to the density $\theta$, for example, the
derivative of the cumulant and the logarithmic derivatives of
$\langle \delta \rangle$ and $\langle \delta^2 \rangle$. In Fig.
3(a), we plot the maximum value of these derivatives as a function
of system size on a log-log scale. The results for $1/\nu$ from
these fits are given in Fig. 3(a). Combining these three
estimates, we obtain $\nu=1.33(1)$. Once we know $\nu$, the
exponent $\gamma$ can be determined by scaling the maximum value
of the susceptibility \cite{Ferren}. Our data for $\chi|_{\rm
max}$ are shown in Fig. 3(b). The value obtained for $\gamma$ is
$\gamma=2.36(4)$.

On the other hand, the standard way to extract the exponent ratio
$\beta/\nu$ is to study the scaling behavior of $\langle \delta
\rangle$ at the point of inflection ${\left({\langle \delta
\rangle |}_{\rm inf}\right)}$, i.e., at the point where $d \langle
\delta \rangle / d \theta $ is maximal. The scaling of ${\langle
\delta \rangle |}_{\rm inf}$ is shown in Fig. 3(b). The linear fit
through all data points gives $\beta^{\left({\langle \delta
\rangle |}_{\rm inf}\right)}=0.139(12)$. In the case of $d \langle
\delta \rangle/d \theta |_{\rm max}$ [see Fig. 3(b)], the value
obtained from the fit is $\beta^{\left(d \langle \delta \rangle/d
\theta |_{\rm max}\right)}=0.138(3)$. Combining the two estimates,
we obtain the final value $\beta=0.139(6)$.

The values calculated for $\nu$, $\beta$ and $\gamma$ clearly
indicate that the IN phase transition belongs to the $q=1$ Potts
universality class (ordinary percolation) \cite{Wu}. This finding
is also consistent with the crossing point of the cumulants shown
in Fig. 2, which is in agreement with recent simulations on 2D
site percolation by Vink \cite{VINK}, where a value of $U^*
\approx 0.638$ has been obtained.

The universality observed in this Rapid Communication can be
interpreted by analyzing the connection between the thermal phase
transition (IN phase transition) occurring in the system and the
behavior of the clusters of aligned monomers. In fact, preliminary
results (based on the observation of the adsorbed state at
critical regime) suggest that, at intermediate concentrations, the
appearance of nematic order at critical density is accompanied by
the simultaneous presence of a cluster of aligned monomers
connecting the extremes of the lattice. In these conditions, an
exact correspondence is expected between $(1)$ the nematic order
parameter and the infinite cluster; $(2)$ the susceptibility and
the mean cluster size; and $(3)$ the IN phase transition and the
percolation phase transition. A similar behavior has been observed
for the site-diluted quenched Ising model at the percolation
threshold \cite{Stauffer}.

The results are also consistent with previous research on polymer
systems in 2D such as polymer networks \cite{Wu1} and branched
polymers \cite{Bunde}, which were shown to be in the universality
class of ordinary percolation. Furthermore, the study of linear
segments of size $k$ and $k$-mers of different structures and
forms deposited on 2D regular lattices has demonstrated that the
system, in all cases, belongs to the random percolation
universality class \cite{Cor}.

%%%%%% FIGURA 4 y 5 %%%%%

The scaling behavior can be further tested by plotting $\langle
\delta \rangle L^{\beta/\nu}$ vs $|\epsilon|L^{1/\nu}$, $\chi
L^{-\gamma/\nu}$ vs $\epsilon L^{1/\nu}$, and $U$ vs $|\epsilon
|L^{1/\nu}$ and looking for data collapsing. Using
$\theta_c=0.524$, and the exact values of the critical exponents
of the ordinary percolation $\nu=4/3$, $\beta=5/36$, and
$\gamma=43/18$, we obtain an excellent scaling collapse as it is
shown in Figs. 4 and 5. This study leads to independent controls
and consistency checks of the values of all the critical
exponents.

In summary, we have used Monte Carlo simulations and finite-size
scaling theory to resolve the nature and universality class of the
IN phase transition occurring in a model of self-assembled rigid
rods. The existence of a continuous phase transition was
confirmed. In addition, as was evident from our results, the
self-assembly process affects the universality of the IN
transition. Thus, the accurate determination of the critical
exponents along with the behavior of Binder cumulants revealed
that the universality class of the IN transition changes from 2D
Ising-type for monodisperse rods without self-assembly to $q=1$
Potts-type for self-assembled rods.

\end{document}